\documentstyle[12pt,aaspp4,flushrt,twoside]{article}

\received{}
\accepted{}
\journalid{}{}
\articleid{}{}

\begin{document}

\markboth{D'Cruz, Sarazin, Dubau}{Li-like {\boldmath$^{57}$}Fe Hyperfine emission}
\title{EXCITATION OF THE ${\boldmath 3.071\,{\rm mm}}$ HYPERFINE LINE IN Li-LIKE
{\boldmath$^{57}$}Fe IN ASTROPHYSICAL PLASMAS}

\author{{\sc Noella L. D'Cruz and Craig L. Sarazin}}
\affil{Department of Astronomy, University of Virginia, \\
P.O. Box 3818, Charlottesville, VA 22903-0818; \\
nld2n@virginia.edu,
cls7i@virginia.edu}

\and

\author{\sc{Jacques Dubau}}
\affil{Observatoire de Paris, F 92195 Meudon Principal Cedex, France; \\
dubau@delacroix.obspm.fr}

\centerline{\it Accepted for publication by the Astrophysical Journal, 27 Jan 1998}

\begin{abstract}
As noted first by Sunyaev \& Churazov (1984), the 3.071~mm hyperfine
line from $^{57}$Fe$^{+23}$ might be observable in astrophysical
plasmas.  We assess the atomic processes which might contribute to the
excitation of this line.  The distorted wave approximation was used to
compute the direct electron collision strength between the two
hyperfine sublevels of the ground configuration; it was found to be
small.  Proton collisional excitation was calculated and found to be
negligible.  We determine the rate of line excitation by electron
collisional excitation of more highly excited levels, followed by
radiative cascades.  The branching ratios for hyperfine sublevels for
allowed radiative decays and electron collisional excitation or
de-excitation are derived.  We show that the dominant line excitation
process is electron collisional excitation of the 2p levels followed
by radiative decay, as first suggested by Sunyaev \& Churazov (1984).
We calculate an effective collision strength for excitation of the
hyperfine line, including all of these effects and correcting for
resonances.  Because the hyperfine line is near the peak in the Cosmic
Microwave Background Radiation spectrum, induced radiative processes
are also very important.  The effect of background radiation on the
level populations and line excitation is determined.

We determine the intensity of the hyperfine line from an isothermal,
coronal plasma in collisional ionization equilibrium.
Because of the variation in the ionization fraction of Fe$^{+23}$,
the emissivity peaks at a temperature of about
$1.8 \times 10^7$ K.
We have also derived the hyperfine line luminosity emitted by
a coronal plasma cooling isobarically due to its own radiation.
Comparisons of the hyperfine line to other lines emitted by the
same ion, Fe$^{+23}$, are shown to be useful for deriving the
isotopic fraction of $^{57}$Fe.
We calculate the ratios of the hyperfine line to
the 2s---2p EUV lines at 192 \AA\ and 255 \AA,
and the 2s--3p X-ray doublet at 10.6 \AA.
\end{abstract}


\newpage

\section{INTRODUCTION} \label{sec:intro}

Astrophysically, the most important and well studied atom exhibiting
hyperfine structure is the neutral hydrogen atom. Its hyperfine
transition at 21~cm has been observed for many years both
in emission and absorption in the gas in our Galaxy and other
galaxies, and has provided information on the velocity structure of
these systems (see Dickey \& Lockman 1990 for a review). Besides this
line, hyperfine emission from $^3$He has also been observed in our
Galaxy in planetary nebulae (Balser et al.\ 1997) and H II regions
(Rood et al.\ 1995).  There have also been a number of searches for
the 92~cm hyperfine line in deuterium (e.g., Lubowich,
Anantharamiah, \& Pasachoff 1989).  In this paper, we are concerned
with another atomic hyperfine transition which could possibly be of
astrophysical interest --- the 3.071~mm line (Shabaeva \& Shabaev
1992) in Li-like $^{57}$Fe.  As originally suggested by Sunyaev \&
Churazov (1984), it is potentially observable in a variety of
astrophysical systems containing very diffuse, hot plasma ($T \sim
10^7$ K), including clusters of galaxies, elliptical galaxies, hot
interstellar gas in our own Galaxy, and supernova remnants.  In a
related paper, we calculate the line intensities expected from cooling
flow and non-cooling flow clusters of galaxies (D'Cruz \& Sarazin
1997; hereafter Paper II).

In the hot plasmas which would emit the $^{57}$Fe hyperfine line,
the dominant radiation is X-ray line and continuum emission.
However, observations of the $^{57}$Fe hyperfine line might have several
advantages compared to X-ray line observations of the same plasma.
First, the atmospheric absorption at 3.071~mm is low enough
to allow ground based observations with large radio telescopes.
Second, the spectral resolution of radio detectors vastly exceeds
that of all existing or planned astrophysical X-ray spectrometers.
Thus, observations of the 3.071~mm line could be used to determine
the velocity field (bulk motions, turbulence, and thermal velocities)
in the hot gas in clusters of galaxies and supernova remnants.
Third, the detection of this $^{57}$Fe hyperfine line would allow the
abundance of this isotope to be determined relative to the more
common $^{56}$Fe isotope.
Since these two isotopes are produced by different nuclear processes,
this would provide a powerful constraint of nucleosynthesis in supernova
remnants and on the chemical history of the intracluster gas.
In this regard, the most direct measure of the isotope ratio comes
from the ratio of the radio line from $^{57}$Fe to the EUV and X-ray lines
from the same ion, Fe XXIV.

The 3.071~mm hyperfine line arises from the ground level
(1s$^2$\,2s $^2$S$_{1/2}$) of $^{57}$Fe XXIV.  The nuclear spin
of the $^{57}$Fe isotope is $I = 1/2$, so the total angular momentum
is $F = 0$ or 1.  We denote the lower hyperfine sublevel ($F=0$) with
the subscript $l$, and the upper hyperfine sublevel ($F=1$) with the
subscript $u$.  The statistical weights of the hyperfine sublevels are
$g_l = 1$ and $g_u = 3$.  The rate of spontaneous radiative decay of
the upper sublevel is $A_{ul} = 9.4 \times 10^{-10} \, {\rm s}^{-1}$
(Sunyaev \& Churazov 1984).

In this paper, we evaluate the atomic processes that lead to
the production of the hyperfine line in Li-like $^{57}$Fe.
We consider direct electron collisional excitation (\S~\ref{sec:direct_e}),
proton collisional excitation (\S~\ref{sec:direct_p}),
indirect excitation to higher levels followed by radiative cascades
(\S~\ref{sec:indirect}),
and stimulated radiative processes due to
the Cosmic Microwave Background Radiation (CMBR) field
(\S~\ref{sec:radexc}).
We present new calculations for the rate of direct electron excitation
of the upper hyperfine line (\S~\ref{sec:direct_e}),
and for indirect excitation of the 2p
sublevels in the same ion, followed by cascade
(\S~\ref{sec:indirect_2p}).
We derive the branching ratios for radiative and collisional transitions
between the hyperfine sublevels of $^{57}$Fe~XXIV
(\S\S~\ref{sec:indirect_cascade}, \ref{sec:indirect_branch}).
A total effective collision strength for exciting the line is given
in \S~\ref{sec:indirect_effective}.
The rate of excitation due resonant scattering of UV and X-ray lines
(optical pumping) is discussed in \S~\ref{sec:indirect_optpump}.
The radiative transfer of the line and of the CMBR is discussed in
\S~\ref{sec:transfer}.
The resulting line intensities in a few simple situations are given in
\S~\ref{sec:results}, where they are also compared to the line intensities
for several EUV and X-ray lines from Fe~XXIV.
Finally, the results are summarized in \S~\ref{sec:summary}.
These excitation rates are used to predict the fluxes of this line
from clusters in Paper II.

\section{DIRECT  COLLISIONAL EXCITATION} \label{sec:direct}

\subsection{\em{Direct Electron Collisional Excitation}} \label{sec:direct_e}

The cross-section for direct electron collisional excitation of the
excited hyperfine sublevel is
$\sigma ( E ) = [ \Omega_{lu} / ( g_l E )] \pi a_o^2$, where
$E$ is the colliding electron kinetic energy (in Ryd), $a_o$ is the
Bohr radius, and $\Omega_{lu}$ is the collision strength.
The direct electron collision strength between the hyperfine sublevels was
calculated using the distorted wave approximation code developed by
Eissner \& Seaton (1972).
For the Li-like target wavefunctions, the SUPERSTRUCTURE code was used.
At the temperatures of interest, the excitation energy of the
upper hyperfine sublevel, $\Delta E_{lu}$, is very small compared to
the kinetic energy of the colliding electron, and the collision is
therefore assumed to be elastic.

The reactance matrix elements corresponding to elastic collisions
within the 1s$^2$\,2s configuration in LS coupling have been converted
to hyperfine structure coupling with $2 \times 4$ 6-j symbols.  This
method of recoupling is similar to the method used for fine structure
reactance matrices (e.g., see the JAJOM program [Saraph 1978]).  To
avoid confusion, we use the same notation as Saraph (1978).  Let
$L_i$, $S_i$, and $J_i$ denote the orbital, spin, and total electronic
angular momenta, respectively, of the Li-like target ion.  Let $l_i$
and $s_i = 1/2$ be the orbital and spin angular momenta of the
electronic projectile.  When we include the nuclear spin $I = 1/2$,
the total angular momentum of the target ion becomes $F_i$, where $\bf
F_i = J_i + I$.  We define the following intermediate couplings: $\bf
K_i = J_i +l_i$ and $\bf G_i = F_i + l_i$.  The subscript $i$
corresponds to any collision channel, while we use $j$ for the initial
channel.  During the collision, the interaction responsible for the
transition is the electron-electron electrostatic interaction.  The
conservation relations for this interaction are $\bf L= L_j + l_j =
L_i + l_i$, $\bf S = S_j + s_j = S_i + s_i$, and $p = p_j = p_i$,
where $p$ is the parity of the system.  Thus, the total electronic
angular momentum of the (target + projectile) system, $\bf J = L + S$,
as well as the total angular momentum including the nuclear spin, $\bf
F = I + J$, are conserved. The other quantum numbers necessary to
define uniquely the target state will be denoted by $\Gamma_i$.  Using
Racah algebra and recoupling of the angular momenta, it is possible to
obtain the reactance matrix elements between $F p$ channels from the
ones between $S L p$ channels as
\begin{eqnarray}
& R^{Fp}(\Gamma_i S_i L_i J_i F_i l_i G_i ; \Gamma_j S_j L_j J_j F_j l_j G_j)
= 
\sum\limits_{S L J \atop K_i K_j} C(S L J K_i, S_i L_i J_i F_i l_i G_i) \hfil &
\nonumber \\
& \qquad \times
R^{S L p}(\Gamma_i S_i L_i l_i; \Gamma_j S_j L_j l_j)\, 
C(S L J K_j, S_j L_j J_j F_j l_j G_j) \, , &
\label{eq:reactance}
\end{eqnarray}
where
\begin{eqnarray}
& C(S L J K_i, S_i L_i J_i l_i G_i) =  (-1)^a \, (2K_i+1) & \hfil \nonumber \\
& \qquad \times [(2S+1)(2L+1)(2J_i+1)(2J+1)(2G_i+1)(2F_i+1)]^{1/2} &
\nonumber \\
& \qquad \times 
\left\{\begin{array}{ccc}
	I & J_i  & F_i \\
	l_i  & G_i & K_i
\end{array} \right\}
\left\{\begin{array}{ccc}
	S_i & L_i & J_i \\
	l_i  & K_i & L
\end{array} \right\}
\left\{\begin{array}{ccc}
	I & K_i  & G_i \\
	1/2  & F & J
\end{array} \right\}
\left\{\begin{array}{ccc}
	L & S_i  & K_i \\
	1/2  & J & S
\end{array}\right\} & \, .
\label{eq:col_coupling}
\end{eqnarray}
The phase factor is $a = S_i - J_i + K_i - I - F_i - G_i - S - J$.
The collision strengths are obtained directly from the reactance matrices
(Saraph 1978).

\begin{figure}[htbp]
\vbox to3.0in{\rule{0pt}{3in}}
\includegraphics{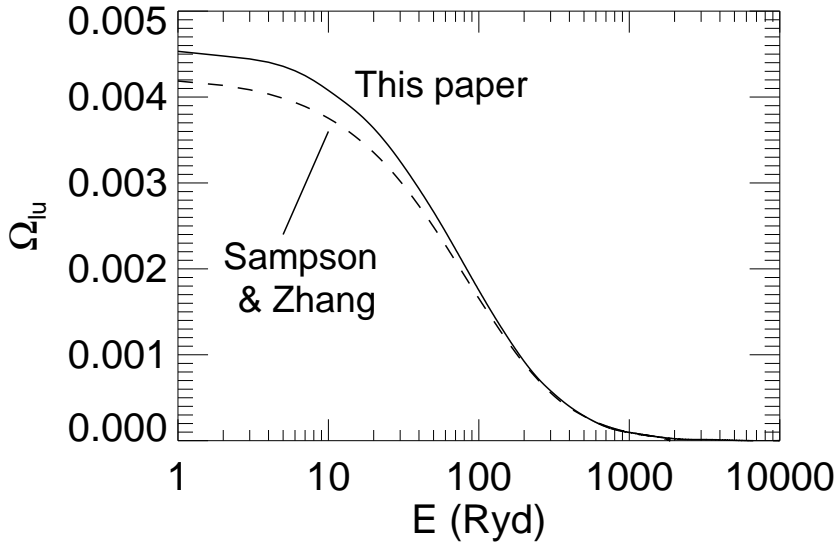}
\figcaption[fig1.ps]{\small  strength for direct electron
excitation of the hyperfine line as a function of the electron
kinetic energy in Rydbergs.
The solid curve gives our result, while the dashed line is from
Sampson \& Zhang (1997).
\label{fig:dircol_e}}
\end{figure}

Figure~\ref{fig:dircol_e} shows the resulting collision strength for
direct electron excitation as function of the kinetic energy of the
electron.  Note that the collision strength is quite small
$(\Omega_{lu} \approx 10^{-3}$) at the energies of interest ($E
\approx 10^2$ Ryd).  Direct collisional excitation is weak, in part,
because the transition is radiatively forbidden, and the forbidden
transition rates drop off rapidly at energies which greatly exceed the
excitation energy.  As we shall see below, the small collision
strength for direct excitation indicates that this is not a very
significant process at temperatures of interest.  While working on
this paper, we received a preprint from Sampson \& Zhang (1997) which
presented relativistic distorted wave calculations of the collision
strength for the same transition.  The Sampson \& Zhang collision
strength is shown as a dashed curve in Figure~\ref{fig:dircol_e}.
These calculations are in good agreement with our calculations; the
Sampson \& Zhang collision strengths are about 7\% lower than ours.

\begin{figure}[htbp]
\vbox to3.0in{\rule{0pt}{3in}}
\includegraphics{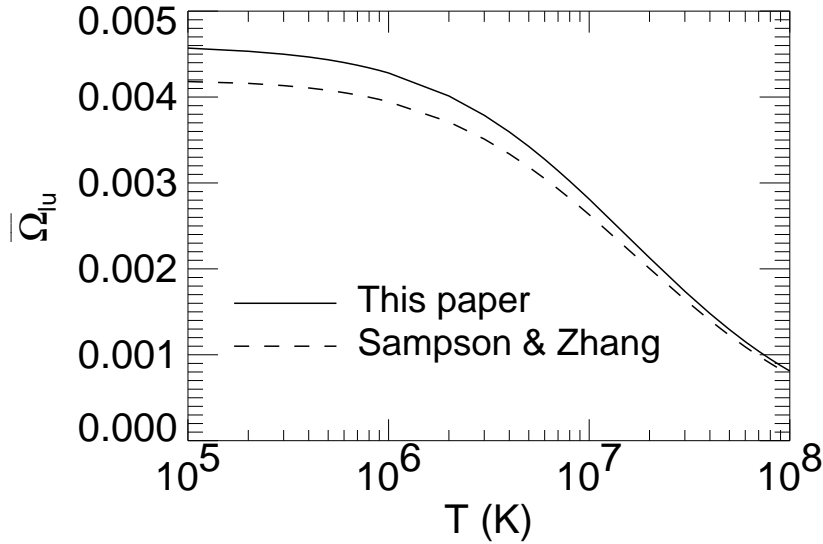}
\figcaption[fig2.ps]{\small Thermally-averaged collision strength for direct
electron excitation of the hyperfine line as a function of temperature.
The solid curve gives our result, while the dashed line is from
Sampson \& Zhang (1997).
\label{fig:dircol_t}}
\end{figure}

The rate of electron collisions which directly excite the
hyperfine line per unit volume is
$n_l n_e q_{lu}^{\rm dir}$, where $n_e$ is
the electron number density and $n_l $ is the number density of
Li-like $^{57}$Fe in the lower hyperfine sublevel.
The rate coefficient is
\begin{equation} 
q_{lu}^{\rm dir}
= \frac{ 8.629\times 10^{-6} \: \bar{\Omega}_{lu}(T)}{g_l \,T^{1/2}}\,
\exp\left( \frac{-\Delta E_{lu}}{k\,T} \right)
\, {\rm cm}^3 \, {\rm s}^{-1} \, .
\label{eq:direct}
\end{equation}
Here,
$\bar{\Omega}_{lu} (T) \equiv \int_0^\infty \Omega_{lu} (E)
\exp ( - E / k T ) \, d (E / k T)$,
is the thermally-averaged direct collision strength.  The value of
$\bar{\Omega}_{lu} (T)$ as a function of temperature is shown in
Figure~\ref{fig:dircol_t}, which is based on the values in
Figure~\ref{fig:dircol_e}, for both our calculations and the Sampson
\& Zhang work. The rate coefficient for de-excitation is related to
equation~(\ref{eq:direct}) by detailed balance.

\subsection{Direct Proton Collisional Excitation} \label{sec:direct_p}

Often, proton collisions contribute significantly to the excitation of
low-lying energy levels at high temperatures ($k T \gg \Delta E$).
This condition clearly applies to the hyperfine sublevels under
consideration, but not so strongly to other excited levels in the
Fe~XXIV ion.
Protons are not important for exciting higher levels in Fe~XXIV,
because they move slower than electrons, undergo Coulomb repulsion, and
do not participate in exchange reactions.

The theory of proton collisional excitation was first developed by
Seaton (1964),
who adapted the theory by
Alder et al.\ (1956) for nuclear Coulomb excitations.
Seaton (1964) and Bely \& Faucher (1970) used this formalism to calculate
the rate of proton collisional excitation of fine structure levels.
This theory can also
be applied to the excitation of hyperfine transitions.
Kastner (1977)
and Kastner \& Bhatia (1979) give detailed expressions for proton
excitation rates based on the theory of Seaton (1964) and Bely \&
Faucher (1970).

We applied these expressions to calculate the direct proton excitation
of the hyperfine sublevels of the Li-like $^{57}$Fe ion.  According to
this theory, proton excitations of the fine and hyperfine structure
sublevels occur primarily through interactions with the electric
quadrupole moment of the ion.  For the Li-like ions, the outermost
free electron is in the 2s state.  Of course, this state is
spherically symmetric, and as such has no quadrupole moment. Hence,
the dominant term in the rate coefficient for proton excitation is
zero.

The next most important term in proton collisional excitation is due
to the magnetic dipole interaction.
In general, magnetic dipole cross-sections are smaller than electric
quadrupole cross-sections by a factor of about $10^{-5}$
(Alder et al.\ 1956; Bahcall \& Wolf 1968).
To check the importance of magnetic dipole excitation rates due to
protons, we compared the rate coefficients for this process with that
due to electron collisions.
The magnetic dipole excitation rates were
several orders of magnitude smaller than the electron collision rates.
Hence, proton collisions can be neglected.

\section{IINDIRECT COLLISIONAL EXCITATION AND CASCADE}
\label{sec:indirect}

The most important process for exciting the hyperfine line turns out
to be electronic collisional excitation to a more highly excited state,
followed by radiative cascade to the excited hyperfine
sublevel.
We refer to this process as ``indirect collisional excitation.''
The rate coefficient for excitation of the upper hyperfine sublevel due
to excitations of higher levels followed by cascades is given by
\begin{equation}
q_{lu}^{\rm indir}= \frac{8.629 \times 10^{-6}}{g_l \,T^{1/2}}
\sum_{k>u} C(k,u) \, \bar{\Omega}_{lk}(T) \,
\exp\left(\frac{-\Delta E_{lk}}{k\,T}\right)
\, {\rm cm}^3 \, {\rm s}^{-1} \, ,
\label{eq:indir}
\end{equation}
where the cascade matrix $C(k,u)$ gives the probability that
the radiative decay of level $k$ will eventually lead to the
excitation of the level $u$.

In this section, we derive the branching ratios for radiative decays,
calculate in detail indirect excitation of the hyperfine line through the
1s$^2$\,2p configuration, give an approximate calculation of the
rate of indirect excitations through more highly excited levels,
and determine an effective collision strength which gives the total
rate of collisional excitation of the hyperfine line.

\subsection{\em{Radiative Cascades}}
\label{sec:indirect_cascade}

With the exception of the upper hyperfine sublevel in the ground
configuration of $^{57}$Fe XXIV, all of the excited sublevels have allowed
radiative decays.
Here, we give the branching ratios among the allowed decays between the
hyperfine sublevels.
We assume LS coupling.
Consider the allowed decays between the an upper level with
total electron spin $S$, total electron orbital angular momentum $L$,
and total electron angular momentum $J$, and a lower level with
quantum numbers $S'$, $L'$, and $J'$.
Let the nuclear angular momentum be $I$.
Let $A(SLJ,S' L' J' )$ be the radiative decay rate between the two
levels, averaging over the hyperfine structure (e.g., the value
normally given in tables of atomic data).
Let $F$ and $F'$ be two hyperfine sublevels of the upper and lower levels,
respectively.
Assume that the hyperfine splittings are small, so that all
of the hyperfine transitions have essentially the same wavelength.
Then, irreducible tensor analysis can be used to derive
the radiative decay rate between the hyperfine sublevels, which is
given in terms of a 6-j symbol as
\begin{equation}
A ( S L J F , S' L' J' F' ) =
A ( S L J , S' L' J' ) (2 F' + 1) (2 J +1)
\left\{
\begin{array}{ccc}
J' & J  & 1 \\
F  & F' & I
\end{array}
\right\}^2 \, .
\label{eq:rad_decay}
\end{equation}
The nuclear spin of $^{57}$Fe is $I= 1/2$.
The total decay rate for any given hyperfine sublevel to all of the
hyperfine sublevels of a lower level is then
\begin{equation}
\sum_{F'} A ( S L J F , S' L' J' F' ) = A ( S L J , S' L' J' )
\, ,
\label{eq:decay_total}
\end{equation}
where this follows from the orthonormality relation for the 6-j symbols.
Thus, the total rates of decay of all of the upper hyperfine levels are
identical and equal to the rate if the hyperfine structure is not
resolved.

Let $P(n J F, n' J' F')$ be the probability that the excited hyperfine
sublevel $n J F$ decays directly to the hyperfine sublevel $n J' F'$,
rather than any other level.
Here, $n$ and $n'$ represents any other quantum numbers which label
the levels.
Thus,
\begin{equation}
P(n J F, n' J' F') \equiv
\frac{ A ( n J F , n' J' F' )}
{\sum_{n'',J'',F''} A ( n J F , n'' J'' F'' )} \, .
\label{eq:prob_hf}
\end{equation}
If one substitutes equation~(\ref{eq:decay_total}) for the decay rate,
and simplifies the denominator using the orthonormality relation for
6-j symbols, the decay probability becomes
\begin{equation}
P(n J F, n' J' F') =
P(n J , n' J') (2 F' + 1) (2 J +1)
\left\{
\begin{array}{ccc}
J' & J  & 1 \\
F  & F' & I
\end{array}
\right\}^2 \, .
\label{eq:prob_hf2}
\end{equation}
Here, $P(n J , n' J')$ is the decay probability if the fine structure
is not resolved,
\begin{equation}
P(n J, n' J') \equiv
\frac{ A ( n J , n' J')}
{\sum_{n'',J''} A ( n J , n'' J'')} \, .
\label{eq:prob_fs}
\end{equation}

A very useful quantity in describing indirect collisional excitation
and cascades is the cascade matrix $C(n J F, n' J' F')$, which is
defined as the probability that an excitation of the $nJF$ hyperfine
sublevel leads to a population of the $n' J' F'$ sublevel by all possible
radiative cascade routes.
Then, if one starts with the obvious result that $C(n J F, n J F) = 1$,
one can determine the cascade matrix recursively from the relationship
\begin{equation}
C(n J F, n' J' F') = 
\sum_{(n'',J'',F'') > (n',J',F')}^{n,J,F}
C( n J F, n'' J'' F'' ) P( n'' J'' F'' , n' J' F') 
\, ,
\label{eq:cascade}
\end{equation}
where the sum extends to all levels above the $n'J'F'$ sublevel,
up to and including the $nJF$ sublevel.
If the upper level $nJ$ is far above the lower level $n' J'$,
and there are many intervening levels with allowed decays, and
complex cascades dominate the cascade matrix, so that
$C (n J F, n' J' F') \gg P (n J F, n' J' F')$ and
$P (n J F, n' J' F') \ll 1$, then equation~(\ref{eq:cascade}) has
the approximate solution
\begin{equation}
C(n J F, n' J' F') \approx
C(n J , n' J') \,
\frac{(2 F' + 1)}{(2 F + 1)} \,
\frac{(2 J + 1)}{(2 J' + 1)} \, ,
\label{eq:cascade_approx}
\end{equation}
where $C(n J , n' J')$ is the cascade matrix if the hyperfine structure
is not resolved.
Thus, complex cascades lead to the populations of hyperfine sublevels
$F'$ in proportion to their statistical weights.

A similar result occurs if the upper hyperfine sublevels $F$ are populated
in proportion to their statistical weights.
Then, the average decay probability to the lower hyperfine sublevels $F'$
is
\begin{equation}
\langle P ( n J, n' J' F' ) \rangle \equiv
\sum_F \, \frac{(2F + 1)}{(2J + 1)(2 I + 1)} \, P (n J F, n' J' F') = 
\frac{(2F' + 1)}{(2J' + 1)(2 I + 1)} \, P (n J, n' J') \, .
\label{eq:prob_average}
\end{equation}
Thus, the lower hyperfine sublevels $F'$ are again populated in proportion
to their statistical weights.

The only levels with allowed decays to the ground configuration in
Fe XXIV are 1s$^2$\,np $^2$P$_J$ levels with $J = 1/2$ or 3/2.
The decay probabilities for these hyperfine radiative transitions are listed in
Table~\ref{tab:rad_branch}.

%
%
\begin{table*}[tbh]
\caption{Hyperfine Radiative Decay Probabilities $P ( J F , J' F' )$
\label{tab:rad_branch}}
\begin{center}
\begin{tabular}{lccccc}
\tableline
\tableline
&\multicolumn{5}{c}{Upper Level} \cr
\cline{2-6}
&\multicolumn{2}{c}{$J=1/2$} && \multicolumn{2}{c}{$J=3/2$} \cr
\cline{2-3} \cline{5-6}
Lower Level&$F=0$&$F=1$&&$F=1$&$F=2$ \cr
\tableline
$F' = 0$ & 0 & 1/3 && 2/3 & 0 \cr
$F' = 1$ & 1 & 2/3 && 1/3 & 1 \cr
\tableline
\end{tabular}
\end{center}
\end{table*}

\subsection{Hyperfine Branching Ratios for Electron Collisional
Excitation}
\label{sec:indirect_branch}

Collision strengths are available in the literature for excitations
to excited levels in Fe XXIV,
although these collisional rates average over the hyperfine structure
in $^{57}$Fe XXIV.
Thus, it is not generally necessary to calculate all of the collision
rates anew.
Instead, what is needed are the branching ratios for the collisional
rates among the hyperfine sublevels.
These branching ratios can be determined using irreducible tensor analysis,
assuming that the hyperfine structure has a negligible effect on the
collision rates.
The interaction Hamiltonian between the bound electron
and the colliding free electron can be expanded as a power series
with terms $( r_<^\lambda / r_>^{\lambda + 1} )$, where $\lambda$ is the
index of the series, and $r_<$ and $r_>$ are the smaller and larger of
the radii of the two interacting electrons.
The interactions are further divided into direct interactions
(D) in which the bound and free electrons retain their identities,
and exchange interactions (E) in which they exchange identities.

For an alkali ion, such as the Li-like system, only the valence electron
is strongly affected during the collision.
Configuration interactions are negligible between target terms, and only one
$\lambda$ term contributes to the direct interaction for
transitions from the 2s configuration.
We can therefore simplify the Racah algebra and reduce
the 8 6-j symbols which arise when equation~(\ref{eq:col_coupling})
is squared to only 2 6-j symbols.
For either direct or exchange reactions, the collision strength for a
transition between the hyperfine sublevels is given approximately by
\begin{equation}
\Omega ( F , F' ) = \Omega ( J , J' )
( 2 F + 1 ) ( 2 F' + 1 ) 
\left\{
\begin{array}{ccc}
J' & J  & \lambda \\
F  & F' & I
\end{array} \right\}^2 \, ,
\label{eq:col_branch_hf}
\end{equation}
where $ \Omega ( J , J' ) $ is the collision strength between the fine
structure levels, ignoring the hyperfine structure and nuclear spin.
An equivalent result has been given by Sampson \& Zhang (1997),
based on jj coupling.
If LS coupling is assumed and the fine structure splitting are also
small, then this branching relationship can be extended to include the
fine structure in direct interactions as
\begin{eqnarray}
\Omega^D ( J F, J' F' ) & = & \Omega^D ( L S, L' S) \,
\frac{ ( 2 J + 1 ) ( 2 J' + 1 ) ( 2 F + 1 )( 2 F' + 1 )}
{ ( 2 S + 1 )} \nonumber \\
& & \qquad \times
\left\{
	\begin{array}{ccc}
		L' & L  & \lambda \\
		J  & J' & S
	\end{array}
\right\}^2 \,
\left\{
	\begin{array}{ccc}
		J' & J  & \lambda \\
		F  & F' & I
	\end{array}
\right\}^2 \, .
\label{eq:col_branch_fs}
\end{eqnarray}
Here, $\Omega^D ( L S, L' S)$ is the direct collision strength between the two
terms, ignoring the fine structure.
For direct interactions, the electron spin $S$ is unchanged.

\subsection{\em{Indirect Excitation Through 2p Sublevels}}
\label{sec:indirect_2p}

\begin{figure}[htbp]
\vbox to7.0in{\rule{0pt}{3in}}
\includegraphics{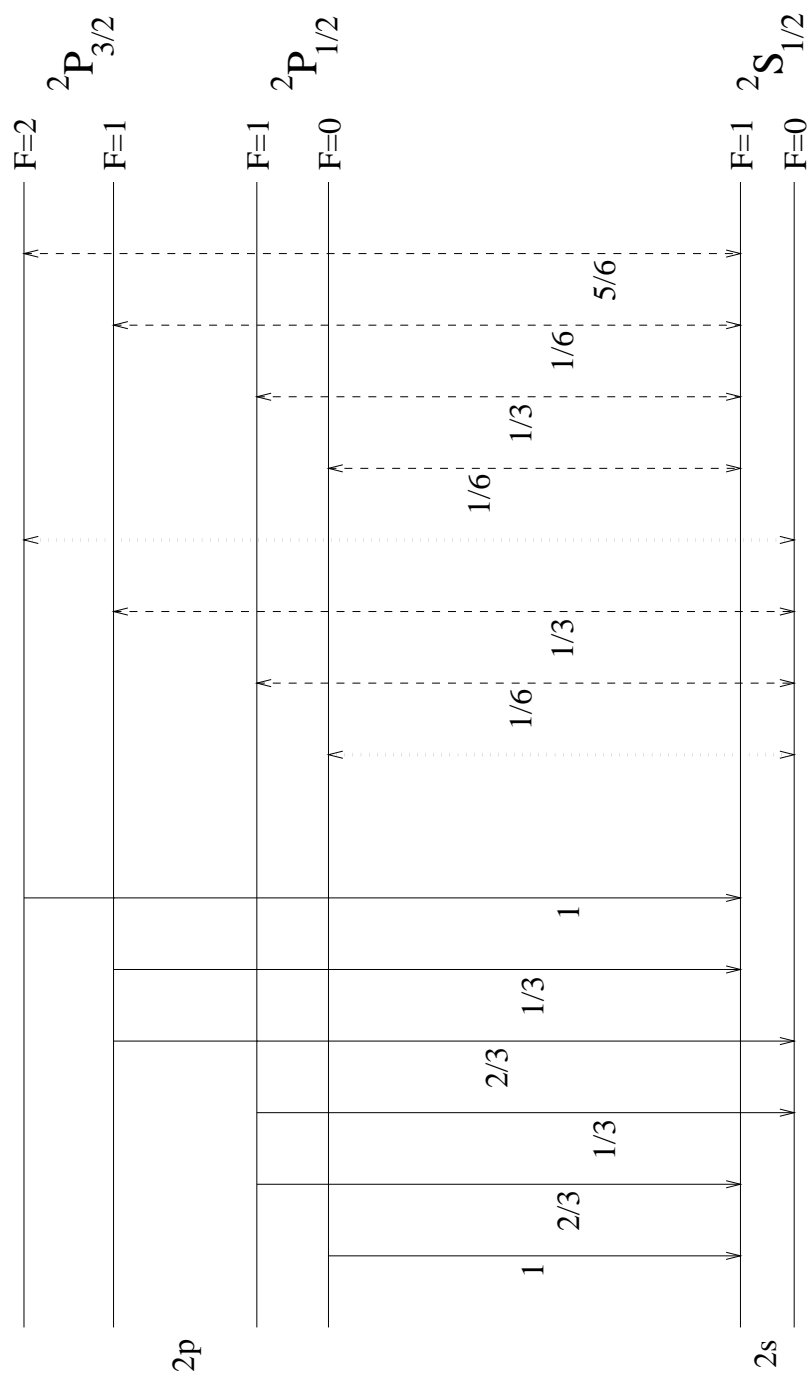}
\figcaption[fig3.ps]{\small A schematic energy level diagram of the 2s and 2p
hyperfine levels for $^{57}$Fe XXIV.
The energies are not shown to scale.
The vertical solid lines on the left show the allowed radiative decays,
and the numbers give the branching ratios as defined in
Table~\protect\ref{tab:rad_branch}.
The vertical lines at the right are the collisional transitions.
The dashed lines give the allowed transitions, and the numbers give
the branching ratios as defined in Table~\protect\ref{tab:col_branch}.
The dotted lines are the weaker forbidden transitions.
\label{fig:levels}}
\end{figure}

In the compilation of collision strengths for Fe XXIV in Gallagher \&
Pradhan (1985), the largest excitation rates from the 2s ground
configuration are to the levels in the low lying 2p configuration.
Thus, we treat the 2p sublevels in detail because their collision
strengths dominate the excitation of the hyperfine line.  These
sublevels are shown schematically in Figure~\ref{fig:levels}.
The allowed radiative decays and collisional excitation and de-excitation
processes are indicated, along with the radiative and collisional
branching ratios.

\begin{figure}[htbp]
\vbox to3.0in{\rule{0pt}{3in}}
\includegraphics{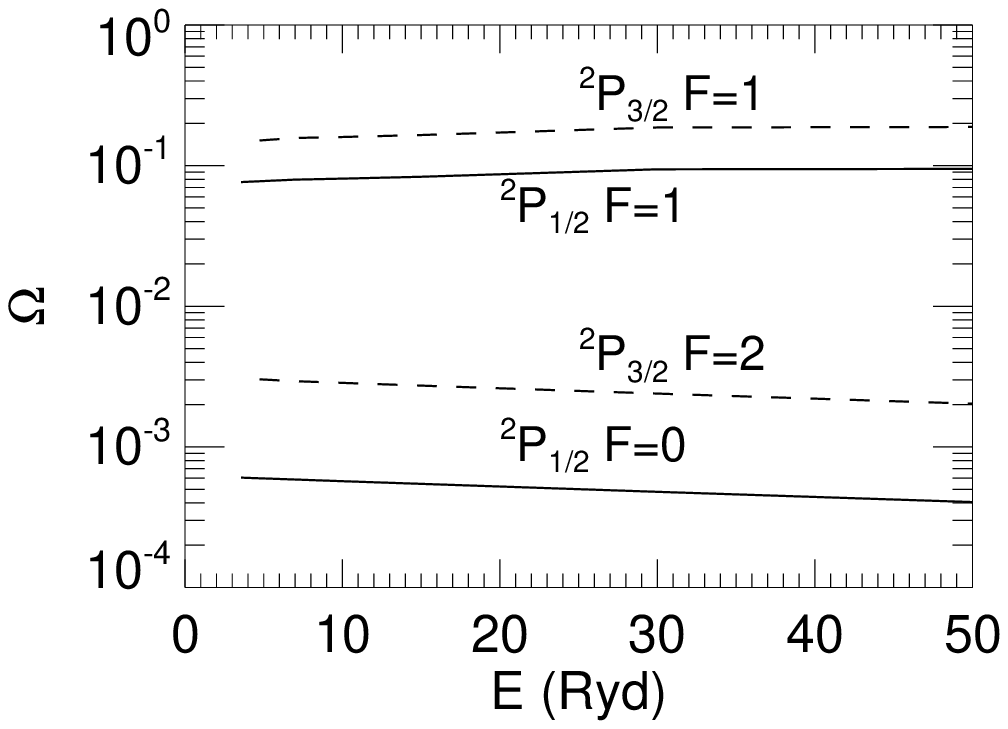}
\figcaption[fig4.ps]{\small Collision strength versus incident electron
energy (in Rydbergs) for electron collisions between the
ground hyperfine 2s $F=0$ levels and the excited
2p hyperfine levels of $^{57}$Fe XXIV.
\label{fig:2pcol_e}}
\end{figure}

For each of the transitions between 2s and 2p hyperfine sublevels, the
collision strength and radiative decay rate (for allowed transitions)
were calculated, using the same distorted wave method and recoupling
scheme discussed in \S~\ref{sec:direct_e}.
In Figure~\ref{fig:2pcol_e}, we illustrate these values by giving the
collision strengths for transitions between the ground hyperfine
2s $F=0$ sublevels and the excited 2p hyperfine sublevels.
The values are given as a function of the incident electron
energy (in Rydbergs) for low energy collision ($E \le 50$ Ryd).
These results illustrate several points about the collisional excitation.

First, even at these low energies, the collisional excitation is dominated
by allowed transitions.
Collisional excitations and de-excitations between the
2s $^2$S$_{1/2}$ $F = 0$ hyperfine sublevel and the
2p $^2$P$_{1/2}$ $F = 0$ or $^2$P$_{3/2}$ $F=2$ hyperfine sublevels
are forbidden.
In Figure~\ref{fig:2pcol_e}, the rates for these forbidden transitions are
at least a factor of 20 smaller than the rates for the allowed
transitions from
the ground 2s $^2$S$_{1/2}$ $F = 0$ hyperfine sublevel to the
2p $^2$P$_{1/2}$ $F = 1$ or $^2$P$_{3/2}$ $F = 1$ sublevels.
When one goes to even higher energies, the allowed transitions dominate
even more strongly.
In collisional ionization equilibrium, the excitation of the
$^{57}$Fe XXIV line should occur mainly at
$ E \approx 10^2$ Ryd or $T \approx 2 \times 10^7$ K
(\S~\ref{sec:results}).
This is considerably greater than the threshold for the
excitation of the 2p $^2$P$_{1/2}$ or 2p $^2$P$_{3/2}$ levels.
Thus, one expects that collisional excitation will be dominated by
radiatively allowed transitions.
As a result of radiative excitation of the hyperfine ground levels
by the cosmic microwave background radiation, the populations of the
two hyperfine ground sublevels are expected to be comparable
(\S~\ref{sec:radexc}).
Now, all of the 2p hyperfine sublevels have allowed transitions from
either one or both of the ground hyperfine sublevels.
Thus, the indirect excitation of the $^{57}$Fe XXIV line through the
2p sublevels will be dominated by allowed electron collisional excitations.

For these allowed electron collisional excitations, the branching ratios
are given by equations~(\ref{eq:col_branch_hf}) or (\ref{eq:col_branch_fs})
with $\lambda = 1$.
Alternatively, the branching ratios can be found by application of the
Van Regemorter (1962) effective Gaunt factor approximation with
the radiative branching ratios in Table~\ref{tab:rad_branch}.
The branching ratios $\Omega(F,F')/\Omega($2s,2p) for the collision
strengths for the allowed 2s to 2p transitions are given in
Table~\ref{tab:col_branch}, where $\Omega($2s,2p) is the total
collision strength between the two terms (ignoring the nuclear spin).
The branching ratios for allowed collisional excitations are also shown
in the left half of Figure~\ref{fig:levels}.

%
%
\begin{table*}[tbh]
\caption{Hyperfine Allowed Collisional Branching Ratio
$\Omega(F,F')/\Omega($2s,2p) \label{tab:col_branch}}
\begin{center}
\begin{tabular}{lccccc}
\tableline
\tableline
&\multicolumn{5}{c}{Upper Level} \cr
\cline{2-6}
&\multicolumn{2}{c}{$J=1/2$} && \multicolumn{2}{c}{$J=3/2$} \cr
\cline{2-3} \cline{5-6}
Lower Level&$F'=0$&$F'=1$&&$F'=1$&$F'=2$ \cr
\tableline
$F = 0$ & 0   & 1/6 && 1/3 & 0   \cr
$F = 1$ & 1/6 & 1/3 && 1/6 & 5/6 \cr
\tableline
\end{tabular}
\end{center}
\end{table*}

Second, the collisional excitations and de-excitations between the
2s $^2$S$_{1/2}$ $F = 0$ hyperfine sublevel and the
2p $^2$P$_{1/2}$ $F = 0$ or $^2$P$_{3/2}$ $F=2$ hyperfine sublevels
are forbidden.
As noted above, the collisions have a significantly reduced rate relative
to the allowed collision.
For forbidden transitions between the singlet 2s $F = 0$
hyperfine sublevel and the other sublevels, the collision strength is
proportional to the statistical weight of the upper sublevel, so
that $\Omega$(2s $F=0$, 2p $F=2)/\Omega$(2s $F=0$, 2p $F=0) = 5$, as shown
in Figures~\ref{fig:2pcol_e}.

Third, the collisional excitation rates based on our calculations of
the collision strength between individual hyperfine sublevels are
in excellent agreement ($\la$7\%) with the values determined by applying
the collisional branching ratios
(Table~\ref{tab:col_branch} or eq.~(\ref{eq:col_branch_hf}) to the
total collision strengths from Hayes (1979).
The values in Hayes are the ones used to calculate EUV and X-ray
lines from the same ion;
as discussed in \S~\ref{sec:results}, these EUV and X-ray lines can
be used with the hyperfine line to determine the abundance of
$^{57}$Fe accurately.
For consistency with with EUV and X-ray analyses of the same
plasma, we have adapted the Hayes (1979) collision strengths
together with the relevant branching ratios.

\subsection{\em{Indirect Excitation Through Higher Levels}}
\label{sec:indirect_higher}

The collision strengths from the ground configuration to more excited
configurations than 2p were obtained from the critical
compilation of Gallagher \& Pradhan (1985).
For most of these transitions, the data come from Hayes (1979). The
rest are from Mann (1983). The standard fitting formulae for collision
strengths given in Clark et al.\ (1982) were used to fit the variation
of each collision strength with temperature.  The collision strength
branching ratios of equation~(\ref{eq:col_branch_hf}) were used to
partition the collisional excitations among the initial and final
hyperfine sublevels.  Equations~(\ref{eq:rad_decay}) and
(\ref{eq:cascade}) were used to determine the cascade matrix from the
hyperfine sublevels.  However, it is worth noting that the collision
strengths to higher levels are all at least 30 times smaller that the
collision strengths to the 2p levels, so their contribution is small.
Sunyaev \& Churazov (1984) first suggested that the dominant line
excitation process for this hyperfine line is electron collisional
excitation of the 2p levels.

Collisional ionization and recombination also make a small
contribution to the overall excitation rate for the hyperfine sublevels.
We assume that these processes populate the hyperfine sublevels in
proportion to their statistical weights.
Note that the cascade matrix among hyperfine sublevels will preserve
such a distribution if it is established in the upper sublevels
(eq.~\ref{eq:prob_average}).

\subsection{\em{Effective Collision Strength and Resonances}}
\label{sec:indirect_effective}

The collision strengths for direct excitation and indirect excitation
of the hyperfine line can be combined to give a single effective collision
strength for the transition.
By combining equations~(\ref{eq:direct}) and (\ref{eq:indir}), one finds
\begin{equation}
\bar{\Omega}_{lu}^{\rm eff} = \bar{\Omega}_{lu} +
\sum_{k>u}
\bar{\Omega}_{lk} (T) \, C(k,u) \exp\left(\frac{-\Delta E_{uk}}{k\,T}\right)
\, .
\label{eq:effective1}
\end{equation}

The distorted wave method used to determine the direct collision strength
did not include the effect of resonances on the collision rate.
Resonances are less important 
when the incident energy greatly exceeds the excitation
energy of the transition.
Thus, we do not expect resonances to strongly affect the
collisional excitation of the hyperfine line.
In collisional excitation through a resonance, the incident free
electron and a bound electron form a bound autoionizing
state, which autoionizes leaving one bound electron in an excited
level.
An approximation to the resonance excitation process is to
treat each resonance as a direct collisional excitation in which the
incoming electron has an energy which is lower than the threshold
$\Delta E_{lk}$ for the excitation
(Seaton 1966;
Petrini 1970;
Mason 1975).
The probabilities for the various decay channels of the autoionizing
state in the resonance are given by the branching ratios for radiative
decay.
These same radiative decay probabilities are
included in the cascade matrix $C(k,u)$.
If the resonance structure is assumed to extend down to the level $u$,
the combined effect of resonances and cascades is given approximately by
equation~(\ref{eq:indir}), with the excitation energy of the highly
excited state replaced by $\Delta E_{lu}$
(Mason 1975).
Thus, we can include resonances approximately by replacing
the effective collision strength in equation~(\ref{eq:effective1}) with
\begin{equation}
\bar{\Omega}_{lu}^{\rm eff} = \bar{\Omega}_{lu} +
\sum_{k>u}
\bar{\Omega}_{lk} (T) \, C(k,u) \, .
\label{eq:effective}
\end{equation}
We use this expression to determine the rate of excitation of the
$^{57}$Fe hyperfine line.
Since most of the excitation is through the 2p levels, and the
excitation energy of these levels is much less than the
temperature at which Fe~XXIV is most abundant, this correction
for resonances is small.

\begin{figure}[htbp]
\vbox to3.0in{\rule{0pt}{3in}}
\includegraphics{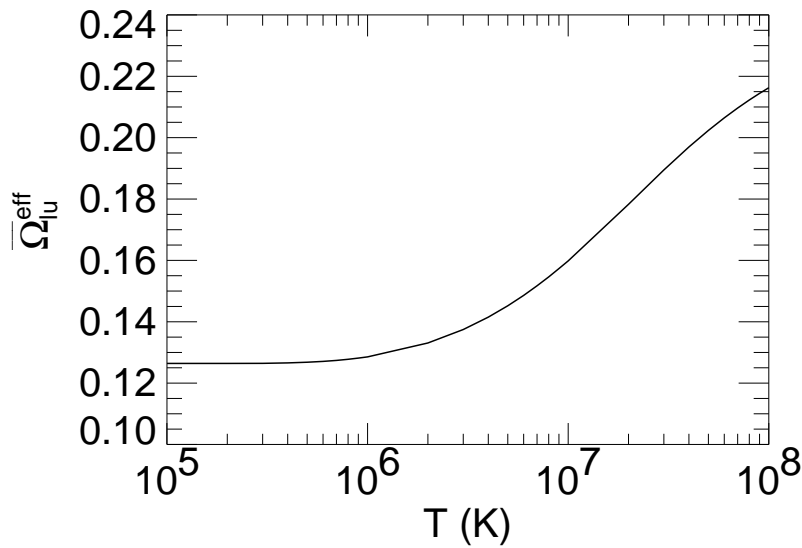}
\figcaption[fig5.ps]{\small Thermally-averaged effective
collision strength for excitation and de-excitation of the $^{57}$Fe
hyperfine line versus temperature.
\label{fig:effective}}
\end{figure}

The resulting values of the effective collision strength as a function
of temperature are shown in Figure~\ref{fig:effective}.
When the electron collision are dominated by direct excitation
and/or allowed collisional excitation followed by direct decay
to the ground hyperfine sublevels, the excitation and de-excitation
rates obey detailed balance, and the same effective collision
strength applies to both excitation and de-excitation.
This is the case for the $^{57}$Fe hyperfine line.

\subsection{\em{Optical Pumping}}
\label{sec:indirect_optpump}

The same indirect electron collisional excitation processes which dominate
the collisional excitation of the $^{57}$Fe hyperfine line also produce
EUV and X-ray photons.
The 2s -- 2p excitations produce EUV lines at 192 \AA\ and 255 \AA,
while the 2s -- 3p excitations make an X-ray doublet at 10.6 \AA.
Photons in these lines, due to the collisional excitation of
Fe$^{+23}$ or due to some unrelated continuum process, can be
absorbed by this ion, and this radiative excitation will normally
be followed by a radiative decay.
This resonant scattering process can also affect the excitation of
the ground hyperfine sublevels through the branching ratios of
the radiative cascades to the sublevels
(Sunyaev \& Churazov 1984).
We refer to this process as ``optical pumping.''

Let $n_l R_{lu} (nSLJ)$ be the rate of excitation per unit volume of ions
in the lower ground hyperfine sublevel to the upper ground hyperfine sublevel
through the resonant scattering of a line whose upper level is denoted
by the quantum numbers $(nSLJ)$.
Let $\bar{J} (nSLJ)$ be the line profile averaged mean intensity of
this EUV or X-ray line, which is defined by
\begin{equation}
\bar{J} (nSLJ) \equiv \int J_\nu (nSLJ) \, \phi(\nu) \, d\nu \, .
\label{eq:meanint}
\end{equation}
Here, $\phi(\nu)$ is the line profile function, which is normalized so
that its integral over all frequencies is unity.
The mean intensity of the line, $J_\nu (nSLJ)$, is averaged over all
directions.
Then, it is straightforward to show that
\begin{equation}
R_{lu} (nSLJ) =
\left[ \frac{c^2 \bar{J} (nSLJ)}{2 h \nu^3} \right]
\sum_{F} \, \frac{2 F + 1}{2 F_l + 1} \,
\frac{A(F , F_l) A(F,F_u)}{\sum_{F'} A(F , F')} \, ,
\label{eq:optpump}
\end{equation}
where $\nu$ is the central frequency of the line,
$F$ denotes a hyperfine sublevel of the upper level $n S L J$,
and $A(F , F')$ is the radiative decay rate of the upper hyperfine sublevel
to one of the ground hyperfine sublevels, $F' = F_l,F_u$.
If one ignores the small difference in $\bar{J} (nSLJ)$ for the hyperfine
components of the line, the symmetrical occurrence of $F_u$ and $F_l$ in
the radiative decay rate in equation~(\ref{eq:optpump}) would imply the
detailed balance relation
$g_l R_{lu} = g_u R_{ul}$.
When one includes the finite hyperfine splitting of the ground level
and treats the line profile in detail including recoil,
the relationship for an optically thick line
becomes (Field 1959)
\begin{equation}
g_l R_{lu} = g_u R_{ul} \,
\exp\left( \frac{-\Delta E_{lu}}{k\,T} \right) \, .
\label{eq:optpump_det}
\end{equation}

Equation~(\ref{eq:optpump}) can be evaluated easily using equations
(\ref{eq:rad_decay}) and (\ref{eq:decay_total}).
For any of the important resonance lines in Fe XXIV 
(2s---2p 192 \AA\ and 255 \AA, and 2s--3p 10.6 \AA),
one finds that
\begin{equation}
R_{lu} (nSLJ) = \frac{2}{3} \,
\left[ \frac{c^2 \bar{J} (nSLJ)}{2 h \nu^3} \right]
A (nSLJ, {\rm 2s}\ ^2{\rm S}_{1/2} ) \, ,
\label{eq:optpump_num}
\end{equation}
where $A (nSLJ, {\rm 2s}\ ^2{\rm S}_{1/2} )$ is the radiative decay
rate to the ground level ignoring the hyperfine structure.
Thus, the rate of optical pumping is simply related to the total
rate of absorption of the line photons.

If the photons being resonantly scattered originated through
collisional excitation of Fe$^{+23}$, optical pumping can be thought
of as increasing the effective rate of indirect collisional excitation.
Crudely, the average rate of collisional excitation will be increased
by a factor which is of the order of the optical depth of the
line.
Under some circumstances, the lines could be moderately optically
thick (optical depths $\la 10^2$), and optical pumping might be
quite important
(Sunyaev \& Churazov 1984).
Because of radiative transfer, this is a nonlocal process, which depends
on the global structure of the system and on the velocity fields.
As a result, we don't discuss optical pumping any further in this
paper.
In Paper II, we include optical pumping in determining the excitation
of the $^{57}$Fe hyperfine line in clusters of galaxies and cooling flows.

\section{LEVEL POPULATIONS AND RADIATIVE EXCITATION} \label{sec:radexc}

Because the $^{57}$Fe hyperfine line wavelength is near the peak of
the Cosmic Microwave Background Radiation (CMBR), it is important to
include the stimulated radiative processes as well as spontaneous
decay.  The rates of stimulated emission and absorption are given by
$B_{ul} \bar{J}$ and $B_{lu} \bar{J}$, respectively.  The Einstein
coefficients $B_{ul}$ and $B_{lu}$ are related to the rate for
spontaneous decay, $A_{ul}$, by the Einstein relations,
\begin{equation}
B_{ul}=\frac{g_l\,B_{lu}}{g_u}=\frac{c^2}{2h \nu_o^3}\,A_{ul} \, ,
\label{eq:einstein}
\end{equation}
where $ \nu_o = \Delta E_{lu} / h$ is the line-center frequency of the
line.
The quantity $\bar{J}$ is the line profile averaged mean intensity of
the radiation at the $^{57}$Fe hyperfine line
(e.g., eq.~[\ref{eq:meanint}]).

The CMBR is a blackbody at a temperature of $T_R = 2.73$ K.
Since the hyperfine line wavelength is near the peak of the CMBR,
one has to use the Planck function without any approximations.
However, the line profile of the hyperfine line is expected to be
fairly narrow, and the line profile averaged mean intensity should
be given accurately by the blackbody intensity at the line
center, $\bar{J} = B_{\nu_o} ( T_R )$, where $B_\nu (T)$ is
the blackbody spectrum.

The time scales for excitation and de-excitation of the 2s hyperfine
levels are much shorter than the ages or any other interesting time scales
in most astrophysical systems of interest.
Thus, we assume that the hyperfine levels are in statistical equilibrium,
\begin{equation}
n_u \left( A_{ul} + B_{ul} {\bar{J}} + n_e q_{ul} \right) =
n_l \left( B_{lu} {\bar{J}} + n_e q_{lu} \right) \, .
\label{eq:statequil}
\end{equation}
Here, $n_u$ and $n_l$ are the number density of ions in the
upper and lower hyperfine sublevels, respectively.
The total rate coefficient for collisional excitation, $q_{lu}$,
is determined by replacing the collision
strength $\bar{\Omega}_{lu}$ with the effective collision
strength $\bar{\Omega}_{lu}^{\rm eff}$ in
equation~(\ref{eq:direct}).
The effective collision strength is given in equation~(\ref{eq:effective})
and Figure~\ref{fig:effective}.

The collisional excitation and de-excitation rates are related by
detailed balance,
\begin{equation}
q_{lu} = \frac{g_u}{g_l} \, q_{ul} \,
\exp \left(-\frac{ \Delta E_{lu}}{k T} \right) \, .
\label{eq:detailb}
\end{equation}
Similarly, the Einstein relations (eq.~\ref{eq:einstein}) can be
used to substitute for $B_{lu}$ and $B_{ul}$ in terms of $A_{ul}$.
With these simplifications,
equation~(\ref{eq:statequil}) can be solved for the ratio of the
occupancy of the upper and lower hyperfine sublevels:
\begin{equation}
\frac{n_{u}}{n_{l}} =
\frac{g_{u}}{g_{l}} \, \exp \left( - \frac{\Delta E_{lu}}{kT} \right) \,
\left\{
\frac{ 1 +
\left( \frac{A_{ul}}{n_e q_{ul}} \right)
\left( \frac{\bar{J} c^2}{2 h \nu_o^3} \right) \, \exp (+ \Delta E_{lu}/kT)}
{ 1 +
\left( \frac{A_{ul}}{n_e q_{ul}} \right)
\left[ 1 +
\left( \frac{\bar{J} c^2}{2 h \nu_o^3 } \right) \right]} \right\} \, .
\label{eq:ratio}
\end{equation}
The quantity $[(\bar{J} c^2)/(2 h \nu_o^3)$ is the photon occupation number
of the radiation field at the frequency of the line.

The above expression implies that when the electron density is very
high, the ratio of level populations approaches LTE at the electron kinetic
temperature $T$.
Given the very small excitation energy $\Delta E_{lu}$, this implies
that the ratio of level populations approaches that of the statistical
weights, $( n_u / n_l ) = 3$, at high densities.
It is useful to define a critical electron density as
\begin{equation} \label{eq:necrit}
n_{e,cr} \equiv \frac{A_{ul}}{q_{ul}} \,
\left[ 1 + \left( \frac{\bar{J} c^2}{2 h \nu_o^3 } \right) \right]
\approx 17 \left( \frac{T}{1.8 \times 10^7 \, {\rm K}} \right)^{1/2}
\, {\rm cm}^{-3} \, ,
\end{equation}
such that the rates of radiative and collisional
de-excitation are equal at this density.
The numerical value in equation~(\ref{eq:necrit}) assumes that the
radiation field is the CMBR at a temperature of 2.73 K, and
the temperature $1.8 \times 10^7$ K is the value at which the
ionization fraction of Fe$^{+23}$ is maximum in collisional ionization
equilibrium (see Figure~\ref{fig:ionfrac}).
The electron density must be at least this high for the level populations
to approach LTE at the electron kinetic temperature.
In most astrophysical situations where the $^{57}$Fe line would be
produced, the density is considerably less than $n_{e,cr}$.

On the other hand, all astrophysical plasmas are immersed in the CMBR.
In the limit of very low densities, the level populations due to the
CMBR are given by
\begin{equation}
\frac{n_{u}}{n_{l}} =
\frac{g_{u}}{g_{l}} \, \exp \left( - \frac{\Delta E_{lu}}{kT_R} \right)
\approx 0.538 \, .
\label{eq:ratio_rad}
\end{equation}
for the temperature of the CMBR in the nearby (i.e., low redshift)
universe.
Thus, the upper hyperfine level is significantly populated even when
there is no collisional excitation.

Because we are most interested in the excitation of the hyperfine
line under low density conditions, it is useful to define a
small parameter $\epsilon$ as
\begin{equation}
\epsilon \equiv \frac{n_e}{n_{e,cr}} \, .
\label{eq:epsilon}
\end{equation}
Then, in the low density limit, the population is given to first
order in $\epsilon$ as
\begin{equation}
\frac{n_{u}}{n_{l}} \approx
\frac{g_{u}}{g_{l}} \, \exp \left( - \frac{h \nu_o}{kT_R} \right)
\left\{
1 + \epsilon 
\left[ \exp \left( \frac{h \nu_o}{kT_R} \right) - 1 \right]
\right\} \, .
\label{eq:ratio_approx}
\end{equation}

\section{RADIATIVE TRANSFER} \label{sec:transfer}

The emissivity, $j_\nu$, and the
absorption coefficient, $\kappa_\nu$, of the hyperfine line are given by
\begin{equation} \label{eq:emiss}
j_{\nu} =
\frac{h\nu}{4 \pi}\,A_{ul}\,n_u \,\phi(\nu) \, ,
\end{equation}
and
\begin{equation} \label{eq:opacity}
\kappa_{\nu} = \frac{A_{ul} \,c^2\, \phi(\nu)}{8\pi
\nu^{2}}\,g_u\,\frac{n_l}{g_l}
\left( 1 -
\frac{n_u}{g_u}\frac{g_l}{n_l}\right) \, .
\end{equation} 
The source function for the line, $S$, is
\begin{equation} \label{eq:source}
S \equiv \frac{j_\nu}{\kappa_\nu} =
\frac{2 h \nu_o^3}{c^2} \, \frac{n_u}{g_u}\frac{g_l}{n_l}
\left( 1 - \frac{n_u}{g_u}\frac{g_l}{n_l} \right)^{-1} \, .
\end{equation} 
In the low density limit (eq.~\ref{eq:ratio_approx}),
the source function is approximately
\begin{equation} \label{eq:source_approx}
S \approx B_{\nu_o} ( T_R )
\left\{
1 + \epsilon \left[
\exp \left( \frac{h \nu_o}{kT_R} \right) - 1 \right] \right\} \, ,
\end{equation} 
Typically, the intensity of the line will be much less than that
of the CMBR, and we can take $T_R = 2.73$ K.

For the moment, we will assume that an emission region producing the
hyperfine line is homogeneous.
Then, the solution of the radiative transfer equation for the
intensity, $I_\nu$, along any given line-of-sight through the region is
\begin{equation} \label{eq:intensity}
I_\nu = I_\nu^o \exp ( - \tau_\nu ) + S
\left[ 1 - \exp ( - \tau_\nu ) \right] \, ,
\end{equation} 
where $I_\nu^o$ is the incident intensity on the far side of the region.
The optical depth is $\tau_\nu \equiv \int \kappa_\nu ds$, where $s$
is the path length along the line-of-sight.

The directly observable quantity is not the brightness of the line,
but the difference between the line and the background radiation
$I_\nu^o$.
This difference is
\begin{equation}
\Delta I_\nu \equiv ( I_\nu - I_\nu^o ) =
( S - I_\nu^o )
\left[ 1 - \exp ( - \tau_\nu ) \right] \, .
\label{eq:netbrightness}
\end{equation}
In most astrophysical environments, the line will be very optically
thin, $\tau_\nu \ll 1$.
Moreover, the background radiation will most often be the CMBR,
and the line will be much weaker than the CMBR.
Thus, we can take $I_\nu^o = B_\nu ( T_R )$.
All of the $^{57}$Fe$^{+23}$ ions are assumed to be in one of the two
hyperfine levels, so that $n( ^{57}{\rm Fe}^{+23}) = ( n_u + n_l )$.
Then, if the low density expression for the source function
(eq.~\ref{eq:source_approx}) is substituted into
equation~(\ref{eq:netbrightness}) and the exponential of the optical
depth is expanded, the resulting net intensity is
\begin{equation}
\Delta I_\nu = \frac{h \nu_o}{4 \pi} \,
D ( T_R ) \,
\int n_e n( ^{57}{\rm Fe}^{+23}) q_{lu} \phi ( \nu ) ds \, .
\label{eq:netbrightness2}
\end{equation}
The correction factor for radiative excitation of the hyperfine structure
$D(T_R)$ is
\begin{equation}
D ( T_R ) \equiv
\frac{ 1 - \exp ( - h \nu_o / k T_R )}{1 +
\frac{g_u}{g_l} \exp ( - h \nu_o / k T_R )}
= 0.533\ {\rm for} \  T_R = 2.73 \, {\rm K} \, .
\label{eq:rad_corr}
\end{equation}
Thus, the line intensity is reduced by about a factor of two below
that expected if there were no radiative excitation of the levels and
all of the electrons were in the lower hyperfine sublevel $l$.
Note that this factor depends somewhat on the redshift $z$ of the emitter,
since $T_R = 2.73 ( 1 + z )$.

For completeness, we note that the hyperfine line intensity in the
astrophysically less interesting high density limit ($n_e \gg n_{e,cr}$)
is given by
\begin{equation}
\Delta I_\nu = \frac{h \nu_o}{4 \pi} \, \frac{g_u}{g_u + g_l} \, A_{ul}
\int n( ^{57}{\rm Fe}^{+23}) \phi ( \nu ) ds \, ,
\label{eq:netbrightness3}
\end{equation}
which is equivalent to the expression for the intensity of the 21~cm
H~I line.
The fraction of the ions in the upper hyperfine sublevel (the second
factor in eq.~[\ref{eq:netbrightness3}]) is 3/4, as in hydrogen.

\section{RESULTING LINE INTENSITIES} \label{sec:results}

\begin{figure}[htbp]
\vbox to3.0in{\rule{0pt}{3in}}
\includegraphics{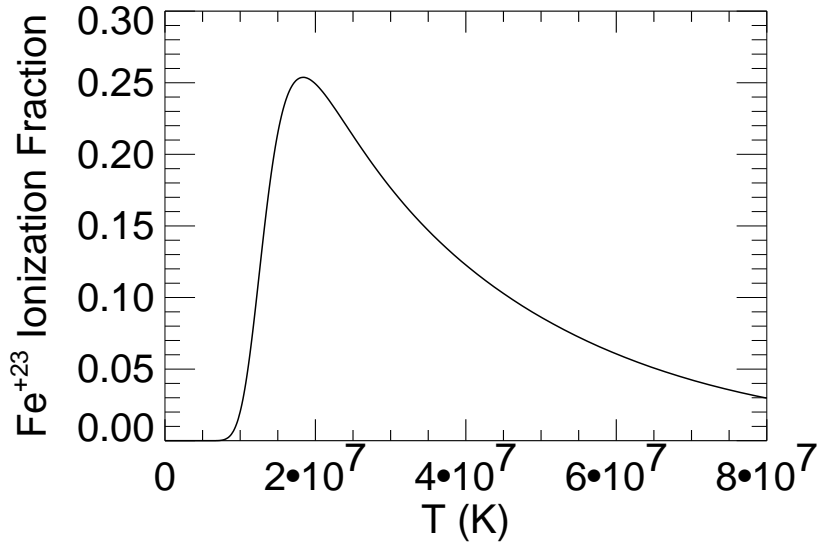}
\figcaption[fig6.ps]{\small The ionization fraction of Fe$^{+23}$ as a
function of temperature.  The ionization fraction peaks at a
temperature of about $1.8 \times 10^7$ K.
\label{fig:ionfrac}}
\end{figure}

We adopted an solar iron abundance of Fe/H = $4.68 \times 10^{-5}$ by
number
(Anders \& Grevesse 1989).
The solar system value of the fraction of iron which is $^{57}$Fe
was taken to be 2.3\%
(V\"olkening \& Papanastassiou 1989).
For consistency with the rates of ionization and emission used in many
analyses of the X-ray emission of hot plasmas, we used the same atomic
physics rates as used in the MEKAL X-ray emission code
(e.g., Kaastra et al.\ 1996), as presented in version 10 of the XSPEC
spectral analysis package.
The MEKAL program includes some recent improvements in the treatment of
the Fe L X-ray lines, including the lines from Fe XXIV
(e.g., Liedahl et al.\ 1995).
The ionization and recombination rates in the MEKAL code are basically
those from
Arnaud \& Rothenflug (1985).
The ionization fraction of Fe$^{+23}$ as a function of temperature is
shown in Figure~\ref{fig:ionfrac}.

\begin{figure}[htbp]
\vbox to3.0in{\rule{0pt}{3in}}
\includegraphics{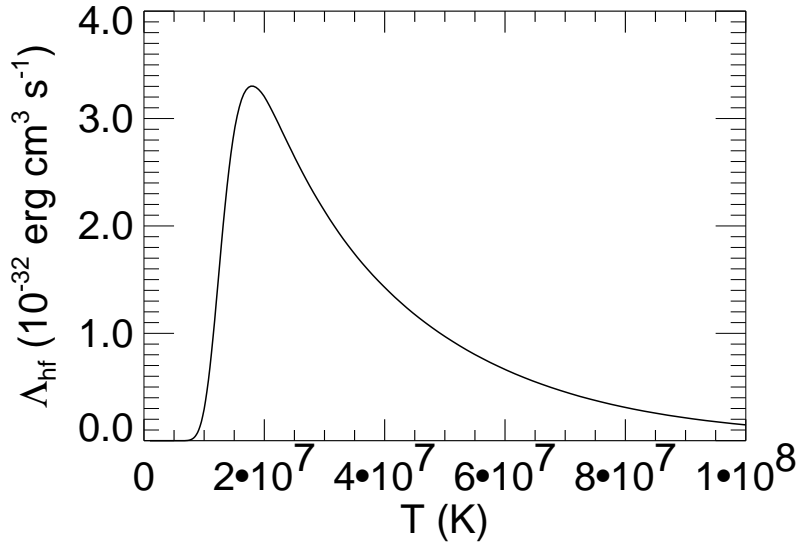}
\figcaption[fig7.ps]{\small The emissivity coefficient $\Lambda_{hf} (T)$
of the $^{57}$Fe$^{+23}$ hyperfine
line as a function of temperature, assuming solar abundances and
collisional ionization equilibrium.
The temperature dependence is mainly due to the ionization fraction of
Fe$^{+23}$ and hence peaks at a temperature of $1.8 \times 10^7$ K.
\label{fig:emissivity}}
\end{figure}

The luminosity of the hyperfine line in the low density limit can
be written as
\begin{equation}
L_{hf} = \Lambda_{hf} (T) \, \int n_p n_e \, dV \, ,
\label{eq:emissivity}
\end{equation}
where the gas is assumed to be isothermal and in collisional ionization
equilibrium.
The integral is the standard integrated emission measure or
emission integral of the plasma, where $n_p$ is the proton number density
and $V$ is the volume of the gas.
The emissivity coefficient $\Lambda_{hf} (T)$ is shown in
Figure~\ref{fig:emissivity}
for solar abundances; it is proportional to the abundance of $^{57}$Fe.
In calculating the values in Figure~\ref{fig:emissivity}, we included
ionization and recombination in the line excitation.
By comparing Figures~\ref{fig:ionfrac} and~\ref{fig:emissivity},
it is clear that the temperature dependence of the emissivity coefficient
is mainly due to the variation of the ionization fraction of Fe$^{+23}$ with
temperature.
The emissivity coefficient peaks around $1.8 \times 10^7$ K.

In the paper which originally suggested that the $^{57}$Fe line might be
of astrophysical interest,
Sunyaev \& Churazov (1984) estimated the excitation rate of
the $^{57}$Fe line along with many other hyperfine lines.
For fixed abundances and ionization fractions, their estimate of
the line excitation rate is about a factor of three higher than
our values.
They estimated that one-half of all excitations of the 2p levels lead
to excitations of the upper hyperfine level.
The actual ratio is slightly lower (4/9 at high energies;
see Tables~\ref{tab:rad_branch} and \ref{tab:col_branch}).
The primary difference would seem to be that their values for the
2s--2p excitation rates are about a factor of three higher than ours.
There is some ambiguity in their paper between the definitions of
excitation and de-excitation rates, and their expression for the
excitation rate may have an extra factor of $(g_u/g_l)$.
This could account for the difference in excitation rates.
When it comes to determining the emissivity of the gas at a given
temperature, their values exceed ours by a larger factor, because of the
assumption of a higher peak ionization fraction for Fe$^{+23}$.
On the other hand, their adopted solar abundance for $^{57}$Fe is
slightly lower than ours.
In any case, it is clear that the intent of the Sunyaev \& Churazov paper
was only to give crude estimates of the intensities of a large number of
hyperfine lines to support the suggestion that they might be observable.

\begin{figure}[htbp]
\vbox to3.0in{\rule{0pt}{3in}}
\includegraphics{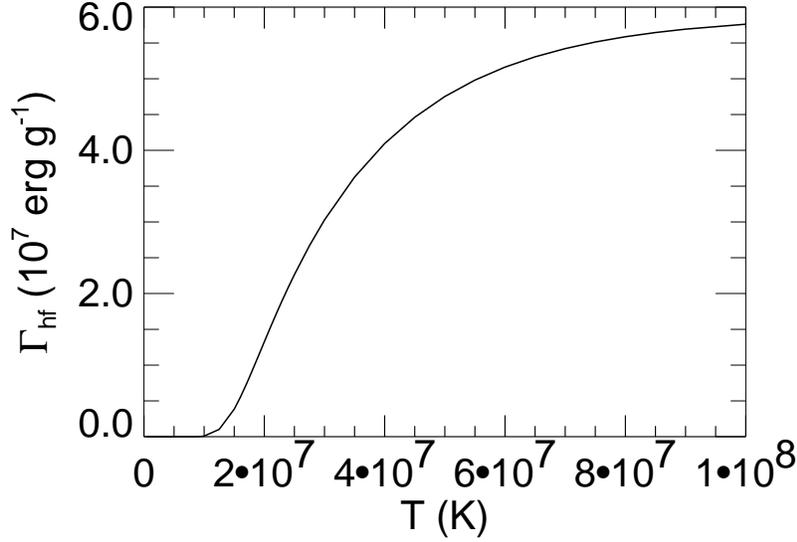}
\figcaption[fig8.ps]{\small The emission of the $^{57}$Fe$^{+23}$ hyperfine
line from a gas cooling isobarically subject to its own radiation.
The function $\Gamma_{hf} (T)$ gives the energy radiated per unit mass
of cooling gas, as a function of the initial temperature of the gas
(eq.~\protect\ref{eq:cooling}).
Solar abundances and collisional ionization equilibrium are assumed.
\label{fig:cooling}}
\end{figure}

We also calculated the emission of the hyperfine line in a plasma
cooling isobarically due to its own radiation, in the low density limit.
The luminosity of the hyperfine line in gas cooling from a temperature
$T$ is given by
(White \& Sarazin 1987)
\begin{equation}
L_{hf} = \dot{M} \,
\left[
\frac{5}{2} \frac{k}{\mu m_H} \,
\int_0^T \frac{\Lambda_{hf}(T')}{\Lambda(T')} \, dT' \right]
= \Gamma_{hf} (T) \dot{M} \, ,
\label{eq:cooling}
\end{equation}
where $\Lambda_{hf}(T)$ is the emissivity coefficient of the hyperfine line,
$\Lambda(T)$ is the coefficient for the total emissivity of the gas
(also sometimes called the cooling function),
$\mu$ is the mass per particle in the gas in units of the
mass of the hydrogen atom $m_H$, and $\dot{M}$ is the rate at which gas
is cooling (in g s$^{-1}$).
The function $\Gamma_{hf} (T)$ is defined by the quantity in brackets
in equation~(\ref{eq:cooling}), and is given in Figure~\ref{fig:cooling}.
$\Gamma_{hf} (T)$ gives the emission per unit mass
of cooling gas, as a function of the initial temperature of the gas.
Again, we used the cooling function derived from the MEKAL code
as presented in XSPEC (version 10).
Solar abundances are used in computing the luminosity.
The value of $\Gamma_{hf}$ rises rapidly with temperature through
the value at which the Fe$^{+23}$ ion is most abundant,
$1.8 \times 10^7$ K.
At temperatures greater than about $7 \times 10^7$ K, the value of
$\Gamma_{hf}$ flattens out, since all gas which starts at higher
temperature passes completely through the temperature range where
the line is emitted strongly.

For the purpose of determining the fraction of iron which is $^{57}$Fe,
it is more useful to compare the strength of the hyperfine line to
other iron lines.
In particular, if the comparison is made to lines produced by
the same ion Fe$^{+23}$, then the derived fractional abundance of $^{57}$Fe
is nearly independent of the ionization structure in the gas.
On the other hand, the absolute flux of the hyperfine line is
very strongly affected by the ionization structure, as is seen
in Fig.~\ref{fig:emissivity}.

\begin{figure}[htbp]
\vbox to3.0in{\rule{0pt}{3in}}
\includegraphics{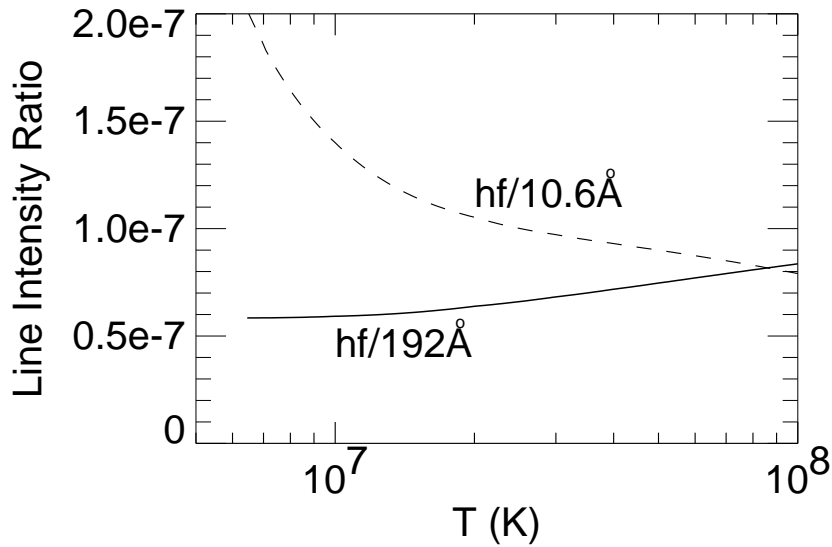}
\figcaption[fig9.ps]{The ratios of the luminosity of the $^{57}$Fe$^{+23}$
hyperfine line to the strongest EUV and X-ray lines from Fe$^{+23}$.
The solid curve is the ratio to the EUV line
2s $^2$S$_{1/2}$ --- 2p $^2$P$_{3/2}$ 192.02 \AA\ line, while
the dashed curve is the ratio to the
2s $^2$S$_{1/2}$ --- 3p $^2$P$_{1/2,3/2}$
10.663, 10.619 \AA\ doublet.
\label{fig:line_ratios}}
\end{figure}

In principle, the best lines for this purpose are the
Fe$^{+23}$ 2s --- 2p lines,
which are 2s $^2$S$_{1/2}$ --- 2p $^2$P$_{3/2}$ 192.02 \AA\ and
2s $^2$S$_{1/2}$ --- 2p $^2$P$_{1/2}$ 255.10 \AA.
These lines are excited by same collisional excitations which we
have found to be the main excitation source for the
Fe$^{+23}$ hyperfine line.
As a result, the ratio of the luminosity of the hyperfine fine
to the luminosity of either of these two extreme ultraviolet (EUV) lines is
nearly independent of the temperature or ionization state of the gas.
To the extent that the excitation of the hyperfine line is dominated
by 2s--2p collisional excitations, the line ratios depend mainly
on the wavelengths of the lines,
the fractional abundance of Fe$^{+23}$,
the radiative correction factor $D(T_R)$ (eq.~\ref{eq:rad_corr}),
and the branching ratios (Table~\ref{tab:col_branch}).
For example, in the limit where 2s---2p excitations dominate the hyperfine
line and the temperature is much greater than the excitation energy of
these transitions, the luminosity ratio to the
2s $^2$S$_{1/2}$ --- 2p $^2$P$_{3/2}$ 192.02 \AA\ line is
\begin{equation}
\frac{L(hf)}{L(192\,{\rm \AA})} \approx \frac{2}{3} \, D ( T_R ) \,
\left( \frac{192\,{\rm \AA}}{3.071\,{\rm mm}} \right) \,
\left[ \frac{n( ^{57}{\rm Fe})}{n( {\rm Fe})} \right]
\approx
2.2 \times 10^{-6}
\left[ \frac{n( ^{57}{\rm Fe})}{n( {\rm Fe})} \right] \, .
\label{eq:192A}
\end{equation}
The term in brackets is the relative abundance of $^{57}$Fe. 
 The equivalent result for the other EUV line (2s $^2$S$_{1/2}$
--- 2p $^2$P$_{1/2}$ 255.10 \AA) is
\begin{equation}
\frac{L(hf)}{L(255\,{\rm \AA})} \approx \frac{4}{3} \, D ( T_R ) \,
\left( \frac{255\,{\rm \AA}}{3.071\,{\rm mm}} \right) \,
\left[ \frac{n( ^{57}{\rm Fe})}{n( {\rm Fe})} \right]
\approx
5.9 \times 10^{-6}
\left[ \frac{n( ^{57}{\rm Fe})}{n( {\rm Fe})} \right] \, .
\label{eq:255A}
\end{equation}
In Figure~\ref{fig:line_ratios}, the $L(hf)/L(192\,{\rm
\AA})$ line ratio is given from a more detailed calculation including
all excitation processes for both lines, and assuming the solar
isotope fraction of $^{57}$Fe of 2.3\%. 
The 192 \AA\ line was calculated using the MEKAL code.
Other excitation processes produce a slight temperature variation
and increase the line ratio by about 20\% above the analytical estimate
at about $120\,{\rm Ryd}$.
As the temperature increases above $2 \times 10^7$ K, the ionization
fraction of Fe$^{+23}$ decreases (Figure~\ref{fig:ionfrac}),
and recombination from Fe$^{+24}$ to Fe$^{+23}$ becomes an important
excitation process for both the hyperfine line and the 192 \AA\ line.
This causes the increase in the line ratio at high temperatures as
seen in Figure~\ref{fig:line_ratios}.

One negative feature of the 2s---2p EUV lines is that they are in
a part of the spectrum which is difficult to observe, partly because of
Galactic absorption.
Moreover, their fluxes might be uncertain even if detected because of
the correction for absorption.
The 2s---3p X-ray lines from Fe$^{+23}$ are easier to observe, and less
subject to absorption.
An alternative standard of comparison might be the allowed doublet
of 2s--3p lines,
2s $^2$S$_{1/2}$ --- 3p $^2$P$_{3/2}$ 10.619 \AA\ and
2s $^2$S$_{1/2}$ --- 3p $^2$P$_{1/2}$ 10.663 \AA.
Because this doublet is difficult to resolve, we consider the
luminosity $L(10.6\,{\rm \AA})$ of the sum of the two lines.
The ratio of the hyperfine line to this doublet is also
shown is Figure~\ref{fig:line_ratios}.
The 10.6 \AA\ line was calculated using the MEKAL code.
Due to the higher excitation energy of this doublet and the difference
in the temperature dependence of the collision strengths
exciting the hyperfine line and the 2s---3p X-ray doublet, this line
ratio is much more temperature dependent.

The observability of this line is crucially dependent on its
wavelength. We use a wavelength value of 3.071 mm from the paper by
Shabaeva \& Shabaev (1992). The uncertainty in this number is about
0.15\% (Shabaev, Shabaeva \& Tupitsyn 1995), which translates into
a velocity uncertainty of about 450 km s$^{-1}$.
This is larger than the bandwidth of many radio spectrometers
in this wavelength region.
On the other hand, the widths of the $^{57}$Fe lines expected in astrophysical
environments may well be even broader (e.g., Paper II).
Hence undertaking observations of this line will be difficult unless a
spectrometer with a large bandwidth is used.
We will discuss the detectability of this line further in Paper II.

\section{SUMMARY} \label{sec:summary}
 
Sunyaev \& Churazov (1984) first showed that the 3.071~mm hyperfine
line from Li-like $^{57}$Fe might be observable in astrophysical
plasmas.
We have assessed the various atomic processes that can contribute
to the excitation of this line.
We calculated the rate of direct electron collisional excitation of
the upper hyperfine level, and found it to be small.
The rate of proton collisional excitation was shown to be negligible.
We calculate the rate of excitation of the line by electron collisional
excitation of more highly excited states, followed by radiative cascade.
We derive the branching ratios for allowed radiative decays to different
hyperfine sublevels, and the resulting cascade matrix.
We also derive branching ratios for the electron collisional excitation
of hyperfine sublevels for direct and exchange interactions.

As originally suggested by Sunyaev \& Churazov (1984), the dominant process
in the excitation of the 3.071~mm hyperfine line is electron collisional
excitation of the 2p levels, followed by cascade.
We calculate the excitation rates for this process, and show that they
are dominated by radiatively allowed transitions, for which the collision
strengths can be derived easily from existing calculations of the collision
strengths and the branching ratios.
We derive an effective collision strength for exciting the hyperfine line
which includes the direct collisional excitation, the excitation of the
2p levels and higher levels, cascades, and a correction for resonances.

At the wavelength of the 3.071~mm hyperfine line, induced radiative
transitions due to the Cosmic Microwave Background Radiation (and possibly,
other microwave radiation sources) are important.
We calculate the effect of the CMBR on the level populations and line
excitation.
We determine the net intensity of the line above the background radiation.
The plasmas that radiate the hyperfine line will most likely have electron
densities that are much lower than the electron densities at which
radiative decays and collisional excitation are balanced.
Thus, we derive expressions for the net line intensity in this
low density limit.

We determine the intensity of the hyperfine line from an isothermal,
coronal plasma in collisional ionization equilibrium with solar
abundances.
Because collisional excitation of the line varies slowly with
is mainly due to the variation in the ionization fraction of Fe$^{+23}$.
For a given emission measure, the line intensity is maximum at a
temperature of about $1.8 \times 10^7$ K.
Our emission rates are somewhat smaller than the estimates given by
Sunyaev \& Churazov (1984).
We have also derived the hyperfine line luminosity emitted by
a coronal plasma cooling isobarically due to its own radiation.

Because of the strong dependence of the hyperfine line emissivity
on the ionization state of the gas, the isotopic abundance of
$^{57}$Fe relative to the total iron abundance is best determined
by comparing the hyperfine line to other lines emitted by the
same ion, Fe$^{+23}$.
We suggest that ratios to the 2s---2p EUV lines at
192 \AA\ and 255 \AA\ or the 2s--3p X-ray lines at 10.6 \AA\ be used
to derive isotopic abundances.
We derive these line ratios as a function of temperature.

In Paper II, we will apply these results to predict the properties
of the 3.071~mm $^{57}$Fe$^{+23}$ hyperfine line from cooling flow
and non-cooling flow clusters of galaxies.

\noindent {\bf Acknowledgements}
C. L. S. was supported in part by NASA Astrophysical Theory Program grant
5-3057.
We thank Bob Brown, Eugene Churazov, Dave Frayer, and Rashid Sunyaev for
useful comments.
We thank Doug Sampson for sending up his results in advance of
publication.


\end{document}